\documentclass[twocolumn]{aastex7}

\usepackage{amsmath}
\usepackage{graphicx}
\usepackage{natbib}
\usepackage{threeparttable}
\usepackage{txfonts}
\hypersetup{allcolors=blue}
\newcommand\iona[2]{#1$\;${\scshape{#2}}}

\begin{document}
\title{The X-ray Link Between High Eddington Ratio Dust-Obscured Galaxies (DOGs) and Hot DOGs}

\author[0000-0002-4436-6923]{Fan Zou}
\affiliation{Department of Astronomy, University of Michigan, 1085 S University, Ann Arbor, MI 48109, USA}
\email[show]{fanzou01@gmail.com}

\author[0000-0002-0167-2453]{W. N. Brandt}
\affiliation{Department of Astronomy and Astrophysics, 525 Davey Lab, The Pennsylvania State University, University Park, PA 16802, USA}
\affiliation{Institute for Gravitation and the Cosmos, The Pennsylvania State University, University Park, PA 16802, USA}
\affiliation{Department of Physics, 104 Davey Laboratory, The Pennsylvania State University, University Park, PA 16802, USA}
\email{wnbrandt@gmail.com}

\author[0000-0001-5802-6041]{Elena Gallo}
\affiliation{Department of Astronomy, University of Michigan, 1085 S University, Ann Arbor, MI 48109, USA}
\email{egallo@umich.edu}

\author[0000-0003-0680-9305]{Fabio Vito}
\affiliation{INAF -- Osservatorio di Astrofisica e Scienza dello Spazio di Bologna, Via Gobetti 93/3, 40129 Bologna, Italy}
\email{fabio.vito@inaf.it}

\author[0000-0002-6990-9058]{Zhibo Yu}
\affiliation{Department of Astronomy and Astrophysics, 525 Davey Lab, The Pennsylvania State University, University Park, PA 16802, USA}
\affiliation{Institute for Gravitation and the Cosmos, The Pennsylvania State University, University Park, PA 16802, USA}
\email{zvy5225@psu.edu}

\begin{abstract}
Dust-obscured galaxies (DOGs) with extremely red optical-to-infrared colors are often associated with intense starburst and AGN activity. Studying DOGs can provide insights into the processes that drive the growth of galaxies and their central supermassive black holes. However, the general DOG population is heterogeneous, spanning a wide range of evolutionary stages, and has \mbox{X-ray} obscuring column densities ($N_\mathrm{H}$) covering low-to-high levels. In this work, we focus on seven high Eddington ratio DOGs ($\log \lambda_\mathrm{Edd} \gtrsim -0.5$) to examine their X-ray obscuration properties using new and archival \mbox{X-ray} observations. We confirm that these systems are generally heavily obscured, with 6/7 having $N_\mathrm{H}\gtrsim10^{23}~\mathrm{cm^{-2}}$ and 3/7 having $N_\mathrm{H}\gtrsim10^{24}~\mathrm{cm^{-2}}$. Based on the observed similarity with the rare Hot DOG population, we argue that both high-$\lambda_\mathrm{Edd}$ DOGs and Hot DOGs likely trace the post-merger phase during which AGNs are enshrouded by large columns of dust-rich material.
\end{abstract}
\keywords{\uat{Active galactic nuclei}{16} --- \uat{Galaxy evolution}{594} --- \uat{AGN host galaxies}{2017}}

\section{Introduction}
\label{sec: intro}
Within the coevolution framework of supermassive black holes (SMBHs) and their host galaxies (e.g., \citealt{Sanders88, Hopkins06, Hopkins08, Alexander12}), major mergers of gas-rich galaxies can trigger intense starburst activity and drive material toward the central SMBHs, fueling accretion. The peak activity of both SMBH accretion and star formation occurs during dust-enshrouded, heavily obscured phases, with the (obscured) SMBH accretion approaching the Eddington limit (e.g., \citealt{Narayanan10, Treister10, Lansbury15, Vito18}). Subsequently, radiation-driven outflows from near the central SMBH sweep away the obscuring material (``blow-out" phase; e.g., \citealt{Hopkins10, Fabian12} and references therein), allowing the SMBH to manifest as an unobscured active galactic nucleus (AGN).\par
Hot dust-obscured galaxies (Hot DOGs) and DOGs have been suggested to represent key post-merger evolutionary phases. Both are selected observationally for their extremely red optical and/or infrared (IR) colors \citep{Dey08, Eisenhardt12, Toba16, Toba17, Wu12, Wu18}, potentially representing the peak phase of SMBH accretion, AGN obscuration, and host star formation in the coevolution framework. Hot DOGs are detected by Wide-field Infrared Survey Explorer (WISE) at 12 and $22~\mu\mathrm{m}$, but nearly undetected at 3.4 and $4.6~\mu\mathrm{m}$ \citep{Eisenhardt12, Wu12}. They are a rare population with a sky surface density of approximately one candidate per $30~\mathrm{deg^2}$ and are thought to be primarily powered by deeply buried, massive SMBHs with high Eddington ratios ($\lambda_\mathrm{Edd}$). Their extreme optical/IR colors are thought to result from emission by hot dust, reaching temperatures up to hundreds of K, heated by central SMBHs (e.g., \citealt{Tsai15, Wu18, Li24}).\par
DOGs are typically selected with $24~\mu\mathrm{m}$ fluxes $f_{24~\mu\mathrm{m}}\ge0.3~\mathrm{mJy}$ and $R$-band to $24~\mu\mathrm{m}$ colors $(R-[24])_\mathrm{Vega}\ge14$ \citep{Dey08}. \citet{Toba16} additionally applied a criterion of $f_{22~\mu\mathrm{m}}\ge3.8~\mathrm{mJy}$ and $i-[22]\ge7$ to select IR-bright DOGs, which tend to have larger AGN contributions. DOGs are generally less extreme, characterized by less-massive SMBHs and larger host-galaxy contributions, and they have smaller intrinsic $2-10$~keV luminosities ($L_\mathrm{X}$) compared to the rarer Hot DOGs. Compared to Hot DOGs, DOGs are more common and representative in the Universe. Numerically, the surface number density of DOGs on the sky can reach $\approx300~\mathrm{deg^{-2}}$ in deep fields (e.g., \citealt{Dey08, Yu24}), but the surface number density of Hot DOGs is four orders of magnitude lower ($\approx0.03~\mathrm{deg^{-2}}$; \citealt{Eisenhardt12, Wu12}).\par
\mbox{X-ray} observations provide unique insights into the nature of (Hot) DOGs. One of the primary advantages of \mbox{X-rays} is their high penetration power, which enables the direct identification of AGNs buried in large column densities ($N_\mathrm{H}$) of obscuring material. Studies have found that Hot DOGs exhibit high $L_\mathrm{X}$ with nearly Compton-thick obscuration \citep{Stern14, Assef16, Assef20, Ricci17, Vito18, Zappacosta18}. Notably, \citet{Vito18} concluded that Hot DOGs occupy a distinct region in the $N_\mathrm{H}-L_\mathrm{X}$ plane, characterized by much higher $N_\mathrm{H}$ than luminous red type~1 quasars, which are thought to be transitioning from a heavily obscured phase to an unobscured blue-quasar phase.\par
However, the X-ray obscuration of DOGs spans a wider $N_\mathrm{H}$ range than that of Hot DOGs \citep{Lanzuisi09, Corral16, Riguccini19, Cristello24, Kayal24, Yu24}. This suggests that the DOG population may be heterogeneous, potentially covering a wide range of evolutionary stages or even sometimes be explained by episodes of star formation, as concluded in \citet{Yu24} based on studying a large sample of well-characterized DOGs. \citet{Yu24} also demonstrated that the general DOG population is not primarily driven by major mergers and that typical AGN-containing DOGs are analogous to extreme type~2 AGNs instead of Hot DOGs. However, \citet{Zou20} argued that high-$\lambda_\mathrm{Edd}$ DOGs naturally tend to be close to the dust-enshrouded phase after major mergers and, like Hot DOGs, are likely heavily obscured in X-rays. We note that previously identified DOG samples rarely reach $N_\mathrm{H}\gtrsim10^{23.5}~\mathrm{cm^{-2}}$, implying that heavily obscured DOGs (i.e., the analogs of Hot DOGs) may be severely underrepresented. The few reported heavily obscured DOGs (e.g, \citealt{Toba20, Zou20, Cristello24}) indeed also have high $\lambda_\mathrm{Edd}$.\par
To further constrain the \mbox{X-ray} obscuration levels of high-$\lambda_\mathrm{Edd}$ DOGs, we compile seven DOGs with known high $\lambda_\mathrm{Edd}$ values that are comparable to those of Hot DOGs (generally having $-0.5\lesssim\log\lambda_\mathrm{Edd}\lesssim0.5$; e.g., \citealt{Li24}) and target four of them with Chandra observations in this work. The detailed sample selection will be presented in Section~\ref{sec: sample}. We will probe if our sources are indeed heavily obscured, as has been seen for Hot DOGs, in Section~\ref{sec: results}. We adopt a flat $\Lambda\mathrm{CDM}$ cosmology with $H_0=70~\mathrm{km~s^{-1}~Mpc^{-1}}$, $\Omega_\Lambda=0.7$, and $\Omega_M=0.3$. For simplicity, we will write SDSS Jhhmmss.ss$\pm$ddmmss.s as Jhhmm$\pm$ddmm in the main text. We will use $L_\mathrm{X}$ and $L_\mathrm{X,obs}$ to represent intrinsic and observed (i.e., without absorption corrections) $2-10$~keV luminosities, respectively, unless otherwise noted. We will always adopt $\mathrm{cm^{-2}}$ as the unit of $N_\mathrm{H}$.

\section{Sample Selection}
\label{sec: sample}
Our sample is primarily drawn from the high-$\lambda_\mathrm{Edd}$ DOG sample in \citet{Zou20}, which was initially from \citet{Toba16} and \citet{Toba17}. \citet{Toba17} systematically selected 36 IR-bright DOGs with $(i-[22])_\mathrm{AB}>7$, $22~\mu\mathrm{m}$ flux densities above 3.8~mJy, and clear [\iona{O}{iii}] from the Sloan Digital Sky Survey. Among them, \citet{Zou20} further selected sources with broad \iona{Mg}{ii} or H$\beta$ lines and targeted 12 DOGs with Cycle~20 Chandra snapshot observations ($\approx3$~ks per source). The goal of selecting broad-line DOGs is to measure their black-hole masses ($M_\mathrm{BH}$), and, consequently, $\lambda_\mathrm{Edd}$, based on their broad-line widths by assuming a virial equilibrium in the broad-line region. Measuring $M_\mathrm{BH}$ also requires intrinsic, deabsorbed AGN optical or UV luminosities (e.g., \citealt{Shen13}). These were derived from SED fitting by recovering the deabsorbed AGN SED components, as has been done in \citet{Zou20}, and such techniques have also been implemented to measure $M_\mathrm{BH}$ values for Hot DOGs (e.g., \citealt{Li24}). \citet{Zou20} showed that these 12 DOGs generally have high $\lambda_\mathrm{Edd}$ and utilized Cycle~20 Chandra observations to examine their basic \mbox{X-ray} obscuration properties. All three targets with $\lambda_\mathrm{Edd}<0.1$ were \mbox{X-ray}-detected, but only 3 out of 9 sources with $\lambda_\mathrm{Edd}>0.1$ were detected. At least at face value, the lower detection rates at higher $\lambda_\mathrm{Edd}$ may suggest that high-$\lambda_\mathrm{Edd}$ DOGs are more obscured. However, the short Chandra exposures in \citet{Zou20} hindered more effective constraints on the \mbox{X-ray} absorption columns of high-$\lambda_\mathrm{Edd}$ DOGs.\par
To further improve the constraints on high-$\lambda_\mathrm{Edd}$ DOGs, we conducted follow-up \mbox{X-ray} observations with longer exposure. Given our available follow-up resources, we focus on the six DOGs with the higher half of $\lambda_\mathrm{Edd}$ in the sample of \citet{Zou20} ($\log\lambda_\mathrm{Edd}\gtrsim-0.5$) as they are more representative of the extreme high-$\lambda_\mathrm{Edd}$ DOG population. As shown in \citet{Li24}, the $\lambda_\mathrm{Edd}$ of Hot DOGs is mainly distributed across $-0.5\lesssim\log\lambda_\mathrm{Edd}\lesssim0.5$ (see their Figures~6 and 7), and thus DOGs below $\log\lambda_\mathrm{Edd}\lesssim-0.5$ in \citet{Zou20} are not formally high-$\lambda_\mathrm{Edd}$ DOGs with $\lambda_\mathrm{Edd}$ comparable to Hot DOGs. Besides, we do not have new observations of these lower-$\lambda_\mathrm{Edd}$ DOGs and thus do not have effective constraints on their $N_\mathrm{H}$. Therefore, the remaining DOGs with lower $\lambda_\mathrm{Edd}$ in \citet{Zou20} will not be discussed in this work.\par
We have obtained a 68~ks follow-up XMM-Newton observation for the X-ray brightest source in the sample, J1324+4501, confirming its heavily obscured nature with $\log N_\mathrm{H}=23.43_{-0.13}^{+0.09}$ \citep{Cristello24}. In this work, we obtained longer Cycle~25 Chandra exposures ($\approx15$~ks per source) for the four fainter sources, where the exposures are set to ensure that, even if these sources are undetected, the corresponding \mbox{X-ray} upper limits are sufficiently tight to confirm the heavily obscured nature. This work presents the new Chandra observations and combines these newly observed sources with the archival ones to provide constraints on the \mbox{X-ray} obscuration levels of high-$\lambda_\mathrm{Edd}$ DOGs.\par
Besides the six high-$\lambda_\mathrm{Edd}$ DOGs from \citet{Zou20}, we also supplement the sample with the single source, J0825+3002, reported in \citet{Toba20}. This DOG was also from the initial IR-bright DOG sample in \citet{Toba16} and observed by XMM-Newton and NuSTAR, showing a Compton-thick $N_\mathrm{H}=1.0_{-0.4}^{+0.8}\times10^{24}~\mathrm{cm^{-2}}$. \citet{Toba20} estimated its $M_\mathrm{BH}$ based on the scaling relation between $M_\mathrm{BH}$ and host-galaxy stellar mass, which places J0825+3002 to also be a high-$\lambda_\mathrm{Edd}$ DOG with $\lambda_\mathrm{Edd}=0.7$. Although this $\lambda_\mathrm{Edd}$ measurement is indirect, we include this source for completeness. We note that explicitly reported high-$\lambda_\mathrm{Edd}$ DOGs are rare because it is generally challenging to confirm their high $\lambda_\mathrm{Edd}$. We include both samples in \citet{Toba20} and \citet{Zou20} to represent the best knowledge we have regarding this population. Removing J0825+3002 out of our sample would not materially affect our overall conclusions in Section~\ref{sec: summary}.\par
We summarize the properties of our sources in Table~\ref{tbl: src}. The \mbox{X-ray} properties are either from the literature (the last three rows of the table) or will be derived in Section~\ref{subsec: nh} (the first four rows of the table). For sources other than J0825+3002, the $M_\mathrm{BH}$ and $\lambda_\mathrm{Edd}$ are from Table~4 in \citet{Zou20} and are all derived from broad \iona{Mg}{ii} lines. For J0825+3002, its $M_\mathrm{BH}$ is based on the scaled host stellar mass, as reported in Section~3.3 of \citet{Toba20}.\par

\begin{table*}
\caption{Source Properties.}
\label{tbl: src}
\centering
\begin{threeparttable}
\begin{tabular}{cccccccc}
\hline
\hline
SDSS Name & $z$ & $\log M_\mathrm{BH}$ & $\log\lambda_\mathrm{Edd}$ & $\log L_\mathrm{X,obs}$ & $\log L_\mathrm{X}$ & $\log N_\mathrm{H}$\\
& & ($M_\odot$) & & ($\mathrm{erg~s^{-1}}$) & ($\mathrm{erg~s^{-1}}$) & ($\mathrm{cm^{-2}}$)\\
(1) & (2) & (3) & (4) & (5) & (6)\\
\hline
J104241.10+245107.0 & 1.026 & $8.80\pm0.15$ & $-0.48\pm0.16$ & $<42.8$ & $44.4_{-0.4}^{+0.3}$ & $>24.0$\\
J121056.92+610551.5 & 0.926 & $7.89\pm0.33$ & $-0.11\pm0.35$ & $42.5_{-0.2}^{+0.3}$ & $43.6_{-0.4}^{+0.8}$ & $23.3_{-0.9}^{+1.6}$\\
J123544.97+482715.4 & 1.023 & $8.20\pm0.12$ & $-0.60\pm0.08$ & $43.4_{-0.1}^{+0.1}$ & $43.8_{-0.1}^{+0.1}$ & $<22.5$\\
J151354.48+145125.2 & 0.882 & $8.15\pm0.12$ & $-0.33\pm0.16$ & $<42.5$ & $44.1_{-0.5}^{+0.3}$ & $>23.9$\\
J082501.48+300257.1$^\star$ & 0.890 & $8.40^a$ & $-0.15^a$ & $44.1$ & $44.6_{-0.2}^{+0.2}$ & $24.0_{-0.2}^{+0.3}$\\
J132440.17+450133.8$^\star$ & 0.774 & $8.27\pm0.40$ & $-0.06\pm0.40$ & $44.11_{-0.04}^{+0.05}$ & $44.71_{-0.12}^{+0.08}$ & $23.43_{-0.13}^{+0.09}$\\
J152504.74+123401.7$^\star$ & 0.851 & $8.36\pm0.23$ & $-0.51\pm0.24$ & $43.3_{-0.2}^{+0.2}$ & $44.3_{-0.3}^{+0.3}$ & $23.2_{-0.3}^{+0.3}$\\
\hline
\hline
\end{tabular}
\begin{tablenotes}
\item
\textit{Notes.} (2) Redshifts. All quoted redshifts are spectroscopic, except for J0825+3002, which only has a photometric redshift. (3) Black-hole masses. (4) Eddington ratios. (5) $2-10$~keV luminosities without absorption corrections. (6) Intrinsic (unabsorbed) $2-10$~keV luminosities. For J1042+2451, J1210+6105, and J1513+1451, these measurements are primarily set by the prior instead of the \mbox{X-ray} data. (7) Intrinsic line-of-sight hydrogen-equivalent column densities. The confidence intervals of the measured values represent 68\% levels, while the limits are at a 95\% level.\\
$^\star$ \mbox{X-ray} measurements for J0825+3002, J1324+4501, and J1525+1234 are from \citet{Toba20}, \citet{Cristello24}, and \citet{Zou20}, respectively.\\
$^a$ The $M_\mathrm{BH}$ of J0825+3002 is estimated based on the scaling relation between $M_\mathrm{BH}$ and the host stellar mass. Therefore, its $M_\mathrm{BH}$ and $\lambda_\mathrm{Edd}$ have nominal uncertainties of $\approx0.5$~dex from the scaling relation.
\end{tablenotes}
\end{threeparttable}
\end{table*}

Although the statistical $\lambda_\mathrm{Edd}$ uncertainties may not allow a firm conclusion on $\log\lambda_\mathrm{Edd}>-0.5$ for individual sources, they are sufficiently small to prove that our sample as a whole can represent high-$\lambda_\mathrm{Edd}$ DOGs -- if adopting 0.5~dex as the nominal uncertainty for J0825+3002's $\log\lambda_\mathrm{Edd}$, the uncertainty of the weighted arithmetic mean $\log\lambda_\mathrm{Edd}$ of our sample would be only 0.06. However, systematic uncertainties are not directly considered here and may be more prominent. The most relevant uncertainty is the viability of using broad lines to infer $M_\mathrm{BH}$ for DOGs. In fact, the origin of the broad lines in such heavily \mbox{X-ray} obscured systems has been under continuous investigation. Relevantly, it is not uncommon for Hot DOGs to show broad lines \citep{Tsai18, Wu18, Finnerty20, Jun20, Li24}. The broad lines of some Hot DOGs are suggested to be potentially from outflows instead of the broad-line region (e.g, \citealt{Finnerty20, Jun20}). However, follow-up studies of a subpopulation of Hot DOGs that show significant excess blue emission \citep{Assef16, Assef20, Assef22} revealed that the blue emission and their broad lines should be the scattered light from central AGNs based on, e.g., multiwavelength and polarimetry observations. \citet{Li24} further argued that, for general Hot DOGs, their broad lines may also be mainly scattered emission, and hence these broad lines can be used to estimate $M_\mathrm{BH}$. For our DOGs specifically, \citet{Zou20} showed that their broad \iona{Mg}{ii} line profiles are generally different from the outflow [\iona{O}{iii}] profiles (see their Section~5.2 for more discussion). Therefore, the broad lines of our sources are highly likely also from scattered emission from the broad-line regions. Nevertheless, an outflow origin cannot be deterministically ruled out, and this is a fundamental caveat for not only our DOGs but also all the Hot DOGs and relevant works in general.

\section{Data and Results}
\label{sec: results}
In this section, we will first present the new Chandra observations (Section~\ref{subsec: chandra}) and $N_\mathrm{H}$ constraints (Section~\ref{subsec: nh}) for four sources in our sample and then combine them with the remaining three archival sources in our sample to discuss the implications for high-$\lambda_\mathrm{Edd}$ DOGs (Section~\ref{subsec: implications}).

\subsection{Chandra Data Reduction}
\label{subsec: chandra}
The four sources with new Chandra Cycle~25 observations also have Cycle~20 observations; we summarize them all in Table~\ref{tbl: obs}. The data were reduced using \texttt{CIAO} 4.16 and \texttt{CALDB} 4.11.2. We first run the \texttt{chandra\_repro} script with the option \texttt{check\_vf\_pha = yes} since our observations were taken in Very Faint mode. We then use the \texttt{fluximage} script to generate images with weights according to a redshifted absorbed power-law model, where the redshift is set to the known value, the intrinsic $N_\mathrm{H}$ is set to $10^{24}~\mathrm{cm^{-2}}$, and the photon index is set to $\Gamma=2$. The weighting parameters are based on prior estimates indicating that our sources are highly obscured. Their exact values are not critical, and altering them does not affect the results. We further use the \texttt{specextract} script to extract \mbox{X-ray} spectra, grouped to at least one count per bin, and response files with $2''$ circular source regions and annulus background regions with inner and outer radii of $10''$ and $40''$ centered at the source position. As shown in Table~\ref{tbl: obs}, there is no apparent count-rate variability between the Cycle 20 and Cycle 25 observations, and we will jointly analyze them hereafter.\par

\begin{table*}
\caption{\mbox{X-ray} Observation Logs.}
\label{tbl: obs}
\centering
\begin{threeparttable}
\begin{tabular}{cccccc}
\hline
\hline
SDSS Name & ObsID & Date & Exposure Time & Counts & Net counts\\
& & (yyyy-mm-dd) & (ks)\\
(1) & (2) & (3) & (4) & (5) & (6)\\
\hline
J104241.10+245107.0 & 21143 & 2018-11-16 & 3.1 & 1 & $<4.7$\\
& 28791 & 2024-05-04 & 15.4 & 0\\
J121056.92+610551.5 & 21141 & 2018-12-24 & 3.1 & 0 & $2.8_{-0.9}^{+2.9}$\\
& 28789 & 2024-04-07 & 15.4 & 3\\
J123544.97+482715.4 & 21140 & 2018-11-06 & 2.9 & 2 & $16.8_{-3.2}^{+5.2}$\\
& 28792 & 2024-11-19 & 14.4 & 15\\
J151354.48+145125.2 & 21145 & 2018-12-26 & 3.1 & 0 & $<3.1$\\
& 28790 & 2024-09-15 & 15.4 & 0\\
\hline
\hline
\end{tabular}
\begin{tablenotes}
\item
\textit{Notes.} (1) SDSS positions in Jhhmmss.ss$\pm$ddmmss.s. (2) Chandra observation IDs. (3) Observation starting date. (4) Exposure time. (5) Total counts within $2''$. (6) Target net counts combining both observations.
\end{tablenotes}
\end{threeparttable}
\end{table*}

To assess whether the sources are detected, we calculate the binomial no-source probability $P_B$ \citep{Broos07, Weisskopf07}. We set a detection threshold of $P_B=0.01$, suitable for sources with known locations. By combining data from Cycle 20 and Cycle 25 observations, we successfully detected J1210+6105 (3 source-region counts) and J1235+4827 (17 source-region counts), with $P_B=1\times10^{-3}$ and $2\times10^{-26}$, respectively; J1042+2451 and J1513+1451 are not  detected.\par
We further constrain the expected net source counts within the source apertures using the method described in Appendix~A of \cite{Weisskopf07}. The 68\% confidence intervals for the counts of detected sources and the 95\% upper limits for undetected sources are presented in Table~\ref{tbl: obs}.

\subsection{$N_\mathrm{H}$ Measurements}
\label{subsec: nh}
\begin{figure}
\centering
\includegraphics[width=\linewidth]{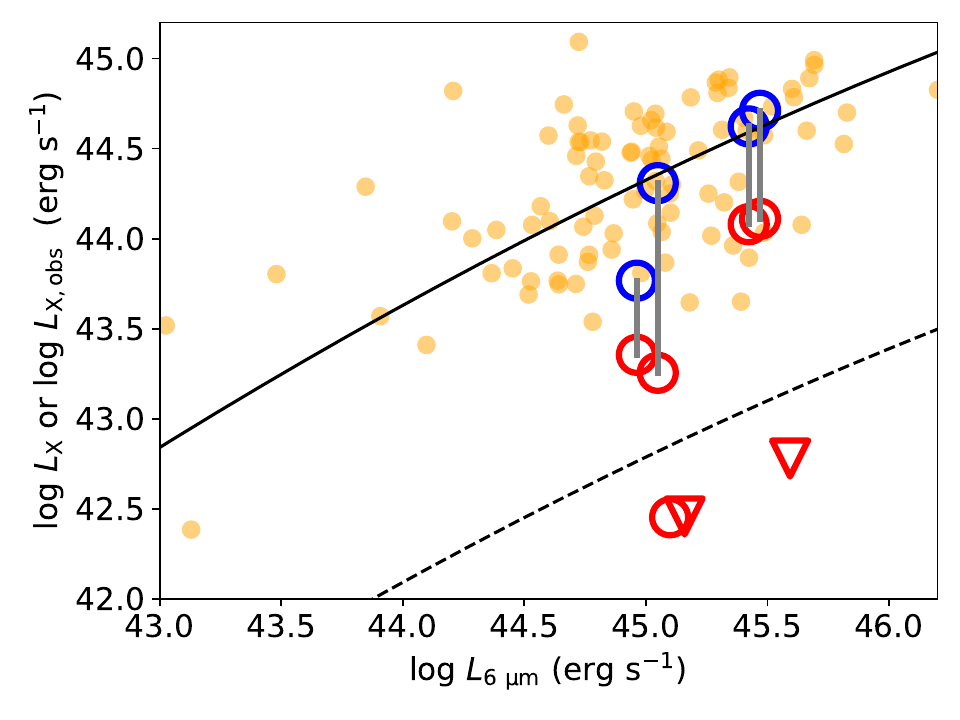}
\caption{$L_\mathrm{X}$ vs. $L_{6~\mu\mathrm{m}}$ for our sample, in larger open circles [downward triangles] representing detected [undetected] sources, and for the DOG sample of \citet{Corral16} and \citet{Yu24}, shown as orange filled circles. The red open circles or triangles represent $L_\mathrm{X,obs}$ for all of our sources, while the blue circles represent $L_\mathrm{X}$ for 4/7 of them; red and blue circles for the same sources are connected with vertical gray lines. All the orange circles are for $L_\mathrm{X}$. These orange circles are mostly not heavily obscured, and thus their intrinsic $L_\mathrm{X}$ and $L_\mathrm{X,obs}$ only differ slightly, with a median offset of 0.2~dex. For comparison, the relation derived by \citet{Stern15} is shown as a solid black line, whereas the dashed black line illustrates the extent to which the relation is suppressed if $N_\mathrm{H}=10^{24}~\mathrm{cm^{-2}}$. The $L_\mathrm{X}$ values of general DOGs and those in our sample are consistent with the AGN $L_\mathrm{X}-L_{6~\mu\mathrm{m}}$ relation. The $L_\mathrm{X,obs}$ values of our sample, on the other hand, are strongly suppressed compared to the relation, possibly indicating heavy \mbox{X-ray} obscuration.}
\label{fig: LxL6um}
\end{figure}

We convert the net counts to $L_\mathrm{X,obs}$ using a power-law model with $\Gamma=1.4$ and the extracted Chandra response files. Figure~\ref{fig: LxL6um} illustrates the comparison between $L_\mathrm{X,obs}$ and the AGN $6~\mu\mathrm{m}$ luminosities, $L_{6~\mu\mathrm{m}}=\nu L_\nu(6~\mu\mathrm{m})$, for the entire sample. The $L_{6~\mu\mathrm{m}}$ values were estimated by \citet{Toba20} and \citet{Zou20} by decomposing the corresponding spectral energy distributions (SEDs). The three archival sources (J0825+3002, J1324+4501, and J1525+1234) have reported intrinsic, deabsorbed $2-10$~keV luminosities $L_\mathrm{X}$ \citep{Toba20, Zou20, Cristello24}. With our new observations, J1235+4827 also has sufficient counts to allow basic $L_\mathrm{X}$ measurements (see the latter part of this section). Therefore, we also plot intrinsic $L_\mathrm{X}$ values for these four sources in the figure. For comparison, we display the expected AGN $L_\mathrm{X}-L_{6~\mu\mathrm{m}}$ relation from \citet{Stern15} and $L_\mathrm{X}$ of typical DOGs reported by \citet{Corral16} and \citet{Yu24}. The goal of presenting intrinsic $L_\mathrm{X}$ instead of $L_\mathrm{X,obs}$ for these typical DOGs is to help evaluate if DOGs follow the AGN $L_\mathrm{X}-L_{6~\mu\mathrm{m}}$ relation. The relation in \citet{Stern15} was empirically constructed for typical AGNs and was shown to be applicable over six orders of magnitude in $L_\mathrm{X}$. Those AGNs with $L_\mathrm{X,obs}$ significantly below the relation may have \mbox{X-ray} emission suppressed by heavy obscuration. For visual guidance, we also plot the expected suppressed $L_\mathrm{X,obs}-L_{6~\mu\mathrm{m}}$ relation when $N_\mathrm{H}=10^{24}~\mathrm{cm^{-2}}$ in the figure, i.e., moving the original $L_\mathrm{X}-L_{6~\mu\mathrm{m}}$ relation downward by 1.5~dex.\par
The $L_\mathrm{X}$ values of typical DOGs generally align with the expected relation. Similarly, for the plotted four DOGs with $L_\mathrm{X}$ measurements in our sample, their $L_\mathrm{X}$ values also agree well with the expected relation. However, all the $L_\mathrm{X,obs}$ of our DOGs fall significantly below this relation, suggesting that their \mbox{X-ray} emission may be heavily suppressed by strong obscuration. Strictly speaking, the suppressed $L_\mathrm{X,obs}$ might also be explained by intrinsic \mbox{X-ray} weakness, but we argue that this is not a preferred explanation. As shown in Figure~\ref{fig: LxL6um}, there is no ambiguity that both general DOGs and 4/7 of our sources are not significantly intrinsically \mbox{X-ray} weak. Regarding the three remaining sources (J1042+2451, J1210+6105, and J1513+1451), their $L_\mathrm{X,obs}$ values are over two dex below the $L_\mathrm{X}-L_{6~\mu\mathrm{m}}$ relation. Even if our high-$\lambda_\mathrm{Edd}$ DOGs can be argued to be different from general DOGs but instead more like Hot DOGs, such a large weakness, if intrinsic, is still not expected by making an analogy to Hot DOGs. Hot DOGs present at most moderate intrinsic \mbox{X-ray} weakness in previous works. \citet{Vito18} found that, on average, Hot DOGs may be slightly systematically below the $L_\mathrm{X}-L_{6~\mu\mathrm{m}}$ relation by 0.2~dex, while \citet{Ricci17} showed a somewhat larger suppression of 0.5~dex.\footnote{Most of the obscuration corrections in \citet{Ricci17} are based on the scaling relation between the optical extinction and $N_\mathrm{H}$, which has a large uncertainty, while the obscuration corrections in \citet{Vito18} are directly based on \mbox{X-ray} data. Therefore, we regard the conclusion in \citet{Vito18} to be more likely.} These differences are far from being sufficient to explain the suppressed $L_\mathrm{X,obs}$ by $>2$~dex for J1042+2451, J1210+6105, and J1513+1451. Therefore, we will pirmarily focus on the heavy-obscuration explanation in the following text.\par

It is important to highlight that all of our sources, regardless of detection status, are unlikely to be solely attributed to starburst activity in the absence of AGN emission, and the AGN signatures exist in both the optical and mid-IR bands. As shown in \citet{Toba17}, these sources have broad \iona{Mg}{ii} lines and have high [\iona{O}{iii}]/H$\beta$ ratios (e.g., see Figure~12 in \citealt{Toba17}) that are characteristics of AGNs. Besides, the mid-IR colors of our sources from the WISE bands, $W1-W2$, are all above 1.4~mag in the Vega system. Such red colors are above the AGN mid-IR color selection criterion, $W1-W2\ge0.8$, in \citet{Stern12} and are expected to be from AGN SEDs because AGNs are much redder than galaxies in the mid-IR. Figure~3 in \citet{Zou20} also showed that strong AGN components are present in the mid-IR.\par
With the limited number of counts, we cannot simultaneously constrain $L_\mathrm{X}$ and $N_\mathrm{H}$ for the newly observed sources. Therefore, we focus on measuring $N_\mathrm{H}$ under the assumption that our sources follow the $L_\mathrm{X}-L_{6~\mu\mathrm{m}}$ relation within a reasonable scatter. The spectral analyses are conducted with \texttt{BXA} \citep{Buchner14} and \texttt{Sherpa} \citep{Siemiginowska24}; the spectra are modeled with \texttt{phabs$\times$(borus01+zphabs$\times$cabs$\times$cutoffpl)}, where \texttt{phabs} represents the Galactic absorption, \texttt{borus01} is the reprocessed spherical torus emission model of \citet{Balokovic18}, \texttt{zphabs$\times$cabs} models the source intrinsic absorption with Compton-scattering losses, and \texttt{cutoffpl} models the \mbox{X-ray} continuum. Since our sources may be in the dust-enshrouded phase, it is appropriate to adopt a covering factor of unity for the \texttt{borus01} model. Note that the \texttt{borus01} component only becomes important when $N_\mathrm{H}\gg10^{24}~\mathrm{cm^{-2}}$, and thus its detailed geometry has little impact on the qualitative inference of the heavy-obscuration nature of our sources. We fix the average $N_\mathrm{H}$ of the torus to $10^{24}~\mathrm{cm^{-2}}$ and use the inferred $N_\mathrm{H}$ values of \texttt{zphabs} and \texttt{cabs} to represent the line-of-sight $N_\mathrm{H}$. We fix the \texttt{cutoffpl} power-law index and cut-off energy to $\Gamma=2$ and 500~keV, respectively. The two free parameters of the model are thus $N_\mathrm{H}$ and $L_\mathrm{X}$. We adopt a flat prior for line-of-sight $\log N_\mathrm{H}$ within $20\le\log N_\mathrm{H}\le26$ and a normal prior for $\log L_\mathrm{X}$ centered at the $L_{6~\mu\mathrm{m}}$-predicted values. The standard deviation ($\approx0.5$~dex) of the $\log L_\mathrm{X}$ prior accounts for the intrinsic scatter of the $L_\mathrm{X}-L_{6~\mu\mathrm{m}}$ relation (0.4~dex) and the $L_{6~\mu\mathrm{m}}$ uncertainties. Note that we are not placing the assumed $L_\mathrm{X}$ exactly on the $L_\mathrm{X}-L_{6~\mu\mathrm{m}}$ relation; instead, the prior dispersion still allows $L_\mathrm{X}$ to deviate away by one dex within the $2\sigma$ range of the prior.\par
We deliberately adopted a physically motivated, weakly informative prior for $L_\mathrm{X}$ because, as mentioned earlier, we would like to measure how much the obscuration level needs to reach to explain the low $L_\mathrm{X,obs}$ \textit{if} our sources follow the $L_\mathrm{X}-L_{6~\mu\mathrm{m}}$ relation within a reasonable scatter. This is a both necessary and reasonable assumption. The necessity is based on the fact that our sources are limited in counts, and thus it is infeasible to simultaneously constrain $L_\mathrm{X}$ and $N_\mathrm{H}$ for, at least, J1042+2451, J1210+6105, and J1513+1451, which have less than three counts each. We can only constrain one of these two parameters (or their limits) after assuming the other one. Our adopted assumption is also reasonable. We have verified that both DOG samples in \citet{Corral16} and \citet{Yu24} are in good consistency with our prior, as can be visually seen in Figure~\ref{fig: LxL6um} -- both samples have a dispersion of 0.38~dex around the expected $L_\mathrm{X}-L_{6~\mu\mathrm{m}}$ relation in \citet{Stern15} and a much smaller median offset. Besides, as shown in Figure~\ref{fig: LxL6um}, the intrinsic $L_\mathrm{X}$ of 4/7 sources in our sample are in good consistency with the $L_\mathrm{X}-L_{6~\mu\mathrm{m}}$ relation. As discussed earlier in this section, previous works also reported that both DOGs and Hot DOGs, either individually or as a sample, are not too far away from the $L_\mathrm{X}-L_{6~\mu\mathrm{m}}$ relation (e.g, \citealt{Ricci17, Vito18, Zappacosta18, Toba20, Cristello24, Yu24}).\par
\begin{figure*}
\centering
\includegraphics[width=\linewidth]{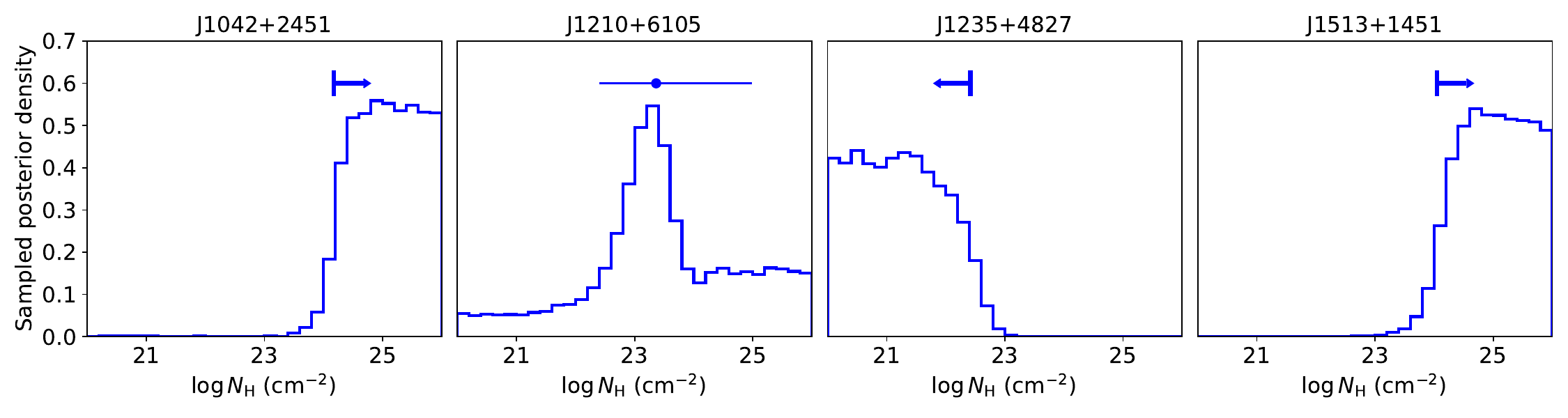}
\caption{The sampled line-of-sight $N_\mathrm{H}$ posteriors of our newly observed sources. The rightward arrows for J1042+2451 and J1513+1451 represent 95\% confidence lower limits, the leftward arrow for J1235+4827 represents a 95\% confidence upper limit, and the point with a horizontal error bar for J1210+6105 represents the posterior median and 68\% interval.}
\label{fig: lognh_sampling}
\end{figure*}

The sampling of the posterior probability distributions is conducted with \texttt{BXA}, which has internally implemented the nested sampling Monte Carlo algorithm in the \texttt{UltraNest} package, as presented in \citet{Buchner21}. We integrate out $L_\mathrm{X}$ and show the sampled marginal line-of-sight $N_\mathrm{H}$ posterior distributions in Figure~\ref{fig: lognh_sampling}. The $N_\mathrm{H}$ posterior of J1235+4827 has an apparent peak at $\log N_\mathrm{H}\approx23.3$. The posteriors of J1042+2451 and J1513+1451 are small at low $N_\mathrm{H}$ and reach plateaus at high $N_\mathrm{H}$, and thus we can constrain their $N_\mathrm{H}$ lower limits. In contrast, J1235+4827 has an $N_\mathrm{H}$ upper limit. Appendix~\ref{append: prior} also shows that the posteriors are generally robust against reasonable changes of the prior. These $N_\mathrm{H}$ measurements are reported in Table~\ref{tbl: src} together with the other three high-$\lambda_\mathrm{Edd}$ DOGs (J0825+3002, J1324+4501, and J1525+1234) from \citet{Toba20}, \citet{Zou20}, and \citet{Cristello24}. The \mbox{X-ray} spectra of these literature-reported DOGs generally have much higher constraining power, and thus their measurements will remain similar no matter if we apply the $\log L_\mathrm{X}$ prior to their data. Table~\ref{tbl: src} indicates that most of our sources are heavily obscured, potentially reaching a Compton-thick level.\par

Table~\ref{tbl: src} presents the inferred intrinsic $L_\mathrm{X}$. The measurements for J1042+2451, J1210+6105, and J1513+1451 are influenced by the prior, which is evident from the large uncertainties in $L_\mathrm{X}$. The intrinsic $L_\mathrm{X}$ values of the remaining sources better reflect the X-ray data constraints.

\subsection{The $N_\mathrm{H}-L_\mathrm{X}$ and $N_\mathrm{H}-\lambda_\mathrm{Edd}$ Planes}
\label{subsec: implications}
\begin{figure}
\centering
\includegraphics[width=\linewidth]{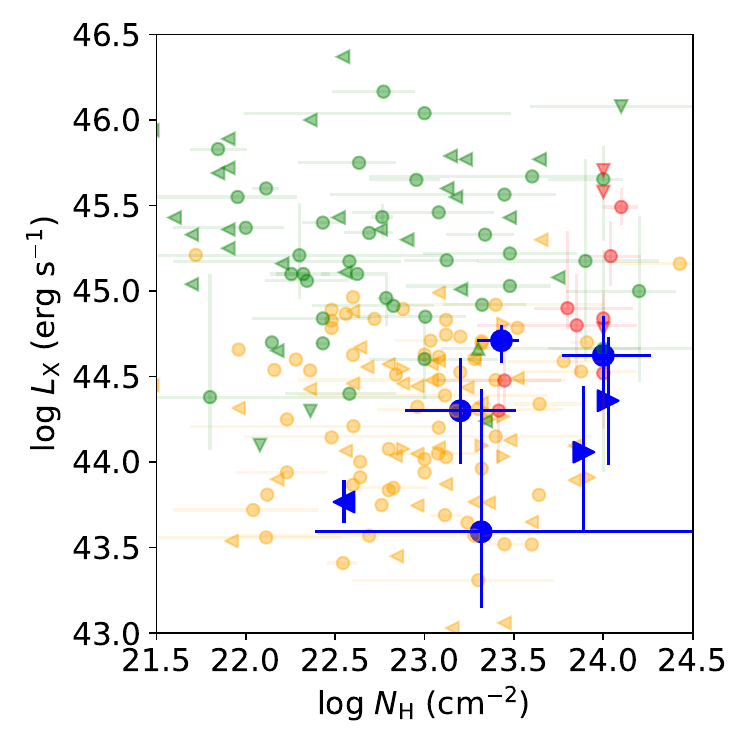}
\caption{The $N_\mathrm{H}-L_\mathrm{X}$ plane. Our sources are plotted as large blue circular points or triangles, where the leftward (rightward) triangles represent $N_\mathrm{H}$ upper (lower) limits. The green, orange, and red points are red type~1 quasars, DOGs, and Hot DOGs, respectively.}
\label{fig: nhlx}
\end{figure}

We display all of our seven sources in the $N_\mathrm{H}-L_\mathrm{X}$ plane in Figure~\ref{fig: nhlx}. For comparison, we also plot red type~1 quasars \citep{Urrutia05, Martocchia17, Mountrichas17, Goulding18, Lansbury20}, Hot DOGs \citep{Stern14, Assef16, Ricci17, Vito18, Zappacosta18}, and DOGs \citep{Lanzuisi09, Corral16, Yu24}. Our sources have $L_\mathrm{X}$ values comparable to other DOGs, but they appear to be generally more obscured, with 6 out of 7 of our sample having $N_\mathrm{H} \gtrsim 10^{23}~\mathrm{cm^{-2}}$. Three of our sources—J0825+3002, J1042+2451, and J1513+1451—reside in the Compton-thick regime, a region that is largely unoccupied by other DOGs. Such a difference also exists between the higher-luminosity Hot DOGs versus red type~1 quasars, suggesting the possible physical link between high-$\lambda_\mathrm{Edd}$ DOGs and Hot DOGs.\par

\begin{figure}
\centering
\includegraphics[width=\linewidth]{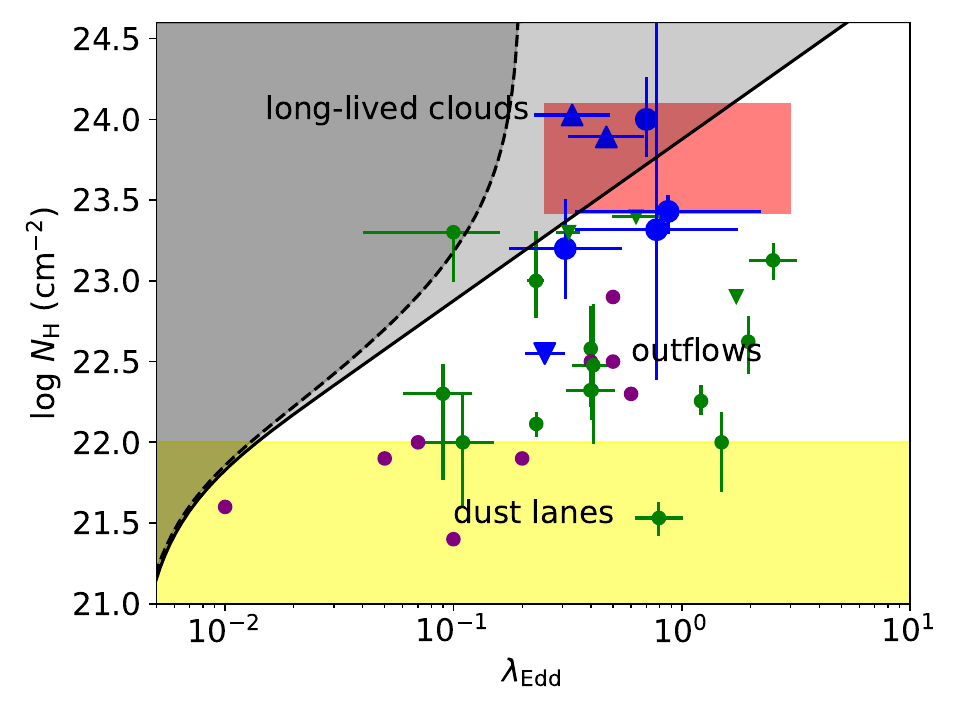}
\caption{The $N_\mathrm{H}-\lambda_\mathrm{Edd}$ plane. Our sources are plotted as large blue circular points or triangles, where the downward (upward) triangles represent $N_\mathrm{H}$ upper (lower) limits. The gray-shaded region is the allowed region for long-lived clouds, and the black solid (dashed) curves represent the boundaries between the allowed region and the outflow region at the single-scattering (radiation-trapping) limit (see \citealt{Ishibashi18} and further descriptions in Section~\ref{subsec: implications}). The yellow region represents obscuration possibly caused by host galaxies \citep{Ricci17b}. The purple points are the sources with strong outflows, the green points represent red type~1 quasars, and the red shaded region is the expected region for Hot DOGs \citep{Ricci17, Vito18, Wu18, Li24}.}
\label{fig: nhledd}
\end{figure}

Furthermore, the $N_\mathrm{H}-\lambda_\mathrm{Edd}$ plane also serves as a useful diagram relevant to AGN-driven outflows (e.g., \citealt{Ishibashi18}). Recall that the classical Eddington limit is defined for pure ionized hydrogen and only includes the electron scattering cross-section. Real astronomical systems, however, usually have dusty gas instead, and dusts have a much higher overall absorption cross-section compared to gas. \citet{Fabian06} introduced the term ``\textit{effective} Eddington limit'' to account for the dust absorption cross section, where AGNs with effective Eddington limits above the unity are expected to generate outflows. As detailed in \citet{Ishibashi18}, the effective Eddington limit can be expressed as a function of $\lambda_\mathrm{Edd}$ and $N_\mathrm{H}$, and the curve of an effective Eddington limit of unity divides the $N_\mathrm{H}-\lambda_\mathrm{Edd}$ plane into a ``forbidden region'' (also known as the outflow region) and an ``allowed region" (e.g., \citealt{Kakkad16, Ricci17b, Ishibashi18, Ishibashi21}). Long-lived, obscuring clouds can only survive in the high-$N_\mathrm{H}$ allowed region, where the obscuring material is sufficiently massive to withstand radiation pressure. Figure~\ref{fig: nhledd} presents the two regions in the $N_\mathrm{H}-\lambda_\mathrm{Edd}$ plane. The boundary between the two regions depends on the IR optical depth of the absorbing material \citep{Ishibashi15}. In the single-scattering limit (i.e., the black solid curve in Figure~\ref{fig: nhledd}), the medium is optically thick to UV but optically thin to IR such that UV photons are absorbed while reemitted IR photons would escape freely. In the radiation-trapping limit (i.e., the black dashed curve in Figure~\ref{fig: nhledd}), the medium is optically thick to both UV and IR, and thus the reemitted IR photons would undergo multiple scatterings and exert larger momentum, leading to a larger forbidden region. Besides the medium in the vicinity of SMBHs, larger-scale dust lanes in their host galaxies may also cause some \mbox{X-ray} absorptions, but the corresponding $N_\mathrm{H}$ should be generally small. We follow \citet{Ricci17b} and \citet{Ishibashi18} and adopt a nominal limit of $N_\mathrm{H}\le10^{22}~\mathrm{cm^{-2}}$ (see Figure~3 in \citealt{Ricci17b}) for the dust lanes (i.e., the yellow region in Figure~\ref{fig: nhledd}).\par

We present our sources in the $N_\mathrm{H}-\lambda_\mathrm{Edd}$ plane in Figure~\ref{fig: nhledd} together with sources with strong outflows \citep{Brusa15, Kakkad16}, red quasars \citep{LaMassa16, LaMassa17, Glikman17, Lansbury20}, and Hot DOGs \citep{Ricci17, Vito18, Wu18, Li24}. Unlike AGNs with strong outflows, our high-$\lambda_\mathrm{Edd}$ DOGs and the Hot DOGs share a similar region in the $N_\mathrm{H}-\lambda_\mathrm{Edd}$ plane and mainly reside at the boundary between the allowed region and the outflow region. Indeed, \citet{Zou20} demonstrated that high-$\lambda_\mathrm{Edd}$ DOGs exhibit only moderate outflows, characterized by outflow-broadened [\iona{O}{iii}] widths of $\lesssim 1000~\mathrm{km~s^{-1}}$, which are narrower compared to sources with strong outflows \citep{Brusa15, Lansbury20}. From this, we conclude that both high-$\lambda_\mathrm{Edd}$ DOGs and Hot DOGs are still in the process of entering the blow-out phase, during which strong AGN outflows start to expel the obscuring materials.\par
We note that J1235+4827, which has the smallest $\lambda_\mathrm{Edd}$ in our sample, is the only source that is not heavily obscured in \mbox{X-rays}. This source thus does not share the similarities with Hot DOGs and may not be in the dust-enshrouded phase after gas-rich galaxy mergers.

\section{Summary}
\label{sec: summary}
In this work, we present new Chandra observations of 4 high-$\lambda_\mathrm{Edd}$ DOGs ($\log\lambda_\mathrm{Edd}\gtrsim-0.5$) and combine them with archival observations for 3 other high-$\lambda_\mathrm{Edd}$ DOGs to probe their \mbox{X-ray} obscuration levels. We found that their $L_\mathrm{X,obs}$ all lie below the expected $L_\mathrm{X}-L_{6~\mu\mathrm{m}}$ relation. Based on the prior knowledge that both DOGs and Hot DOGs are not strongly intrinsically \mbox{X-ray} weak, we argue that the low $L_\mathrm{X,obs}$ is caused by heavy obscuration. We further constrain the $N_\mathrm{H}$ with the inclusion of physically motivated $\log L_\mathrm{X}$ priors constructed based on the $L_\mathrm{X}-L_{6~\mu\mathrm{m}}$ relation. We find that these systems are generally heavily obscured, with 6/7 having $N_\mathrm{H}\gtrsim10^{23}~\mathrm{cm^{-2}}$ and 3/7 having $N_\mathrm{H}\gtrsim10^{24}~\mathrm{cm^{-2}}$.\par
Previous work \citep{Zou20} established that high-$\lambda_\mathrm{Edd}$ DOGs share several important similarities with Hot DOGs: (i) their central SMBHs have high accretion rates; (ii) they are still entering the blow-out phase and do not have sufficiently strong outflows to sweep away large column densities of enshrouded materials; (iii) they are associated to intense host-galaxy starbursts. This work confirms that high-$\lambda_\mathrm{Edd}$ DOGs, similar to Hot DOGs, are also heavily obscured in \mbox{X-rays}, potentially up to a Compton-thick level. These properties are consistent with the peak phase of AGN obscuration and SMBH and galaxy growth after gas-rich galaxy mergers and suggest that high-$\lambda_\mathrm{Edd}$ DOGs and Hot DOGs are at similar evolutionary stages.\par
Nevertheless, as argued in previous works (e.g., \citealt{Yu24}), DOGs may originate from a heterogeneous population that cannot always be explained by the merger-driven coevolution framework. In fact, Section~4.1 in \citet{Yu24} even showed that the fraction of DOGs hosting AGNs is similar to that of the typical matched galaxy population. Although the $\lambda_\mathrm{Edd}$ distribution of DOGs is still unknown, this may indicate that the majority of general DOGs do not have high $\lambda_\mathrm{Edd}$. As Figure~\ref{fig: nhlx} shows, general DOGs span a wide range of $N_\mathrm{H}$ and are often not heavily obscured in \mbox{X-rays}. However, once we only focus on high-$\lambda_\mathrm{Edd}$ DOGs, as represented by our sample, our results indicate that such sources would almost all populate in the heavily obscured regime. Therefore, our findings likely suggest that $\lambda_\mathrm{Edd}$ may be a key factor in distinguishing different types of DOGs. Within the DOG population, only those with high $\lambda_\mathrm{Edd}$ may serve as the less-massive analogs to Hot DOGs.\par
One of the main challenges in characterizing these sources is their low \mbox{X-ray} counts, which are a result of their high $N_\mathrm{H}$. The limited counts may have hindered the \mbox{X-ray} identification of high-$\lambda_\mathrm{Edd}$ DOGs in previous studies, and it is likely that those DOGs in the dust-enshrouded phase are still largely missed. With current X-ray facilities (e.g., Chandra and XMM-Newton), it would require several hundred kiloseconds per source to gather sufficient X-ray photons for detailed spectral analysis, making it a costly endeavor. Future X-ray observatories (e.g., NewAthena and AXIS) should offer much higher throughput, enabling better characterization of the dust-enshrouded, heavily obscured phase of merger-driven events.

\begin{acknowledgments}
We thank the anonymous referee for constructive suggestions and comments. WNB acknowledges support from Chandra X-ray Center grant GO2-23083X and the Penn State Eberly Endowment. FV acknowledges support from the ``INAF Ricerca Fondamentale 2023 -- Large GO'' grant. The Chandra ACIS team Guaranteed Time Observations utilized were selected by the ACIS Instrument Principal Investigator, Gordon P. Garmire, currently of the Huntingdon Institute for X-ray Astronomy, LLC. This paper employs a list of Chandra datasets, obtained by the Chandra X-ray
Observatory, contained in~\dataset[DOI: 10.25574/cdc.417]{https://doi.org/10.25574/cdc.417}.
\end{acknowledgments}

\appendix
\section{Testing Different $\log L_\mathrm{X}$ Priors}
\label{append: prior}
\begin{figure}
\centering
\includegraphics[width=\linewidth]{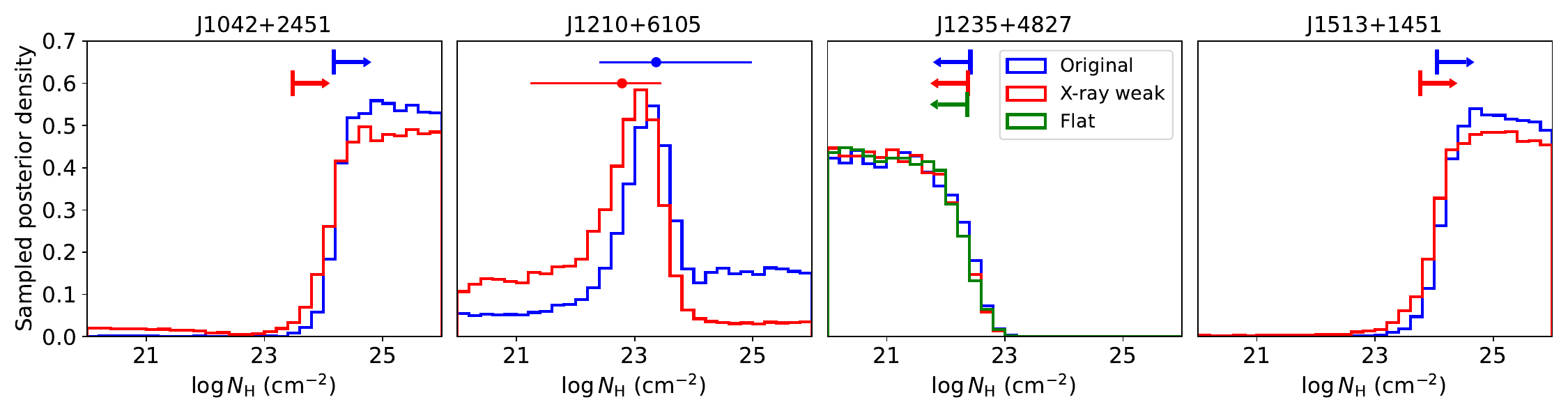}
\caption{Comparison of the $\log N_\mathrm{H}$ posterior under different $\log L_\mathrm{X}$ prior choices. As labeled in the legend of the third panel, the blue histograms and the corresponding arrow limits (95\% confidence) or error bars (68\% confidence) are for the original results, i.e., the same as those in Figure~\ref{fig: lognh_sampling}; the red ones are based on the $\log L_\mathrm{X}$ priors systematically shifted downward by 0.5~dex; and the green one for J1235+4827 represents a flat $\log L_\mathrm{X}$ prior.}
\label{fig: comp_nh_prior}
\end{figure}

The main text assumes the $\log L_\mathrm{X}$ priors to center at the $L_{6~\mu\mathrm{m}}$-predicted values. We also try shifting them downward by 0.5~dex to represent the fact that \citet{Ricci17} found a similar level of intrinsic \mbox{X-ray} weakness for Hot DOGs. This deviation is already at an extreme level because \citet{Vito18} showed that the intrinsic \mbox{X-ray} weakness of Hot DOGs may be milder (only 0.2~dex downward; see Footnote~6), and Figure~\ref{fig: LxL6um} indicates that DOGs are generally not \mbox{X-ray} weak. Therefore, we do not move the priors down further because we want them to be still somewhat physically plausible. We conducted the same analyses as in Section~\ref{subsec: nh} and compare the marginal $\log N_\mathrm{H}$ posterior in Figure~\ref{fig: comp_nh_prior}. The figure shows that, even if this \mbox{X-ray} weak prior is adopted, the $\log N_\mathrm{H}$ posteriors remain largely similar, and the 95\% lower limits of J1042+2451 and J1513+1451 are still above $10^{23}~\mathrm{cm^{-2}}$. Our overall qualitative conclusions in the main text would not be affected.\par

J1235+4827 has many more counts than the other three sources (see Table~\ref{tbl: obs}). Therefore, it is feasible to further relax the assumption on the $\log L_\mathrm{X}$ prior. We tried a flat $\log L_\mathrm{X}$ prior and present the corresponding result in the third panel of Figure~\ref{fig: comp_nh_prior}. The $\log N_\mathrm{H}$ posterior of J1235+4827 almost remains the same regardless of the prior choices, indicating that its inference is already data-dominated. We do not test a flat, noninformative prior for the other three sources because it would not provide any insights. For example, an extremely low $\log L_\mathrm{X}$ (even reaching $-\infty$) would certainly be favoured for J1042+2451 and J1513+1451 because they do not have counts, but this case is physically unlikely. The necessity of a weakly informative prior instead of a noninformative prior for statistical inferences beyond the fully data-dominated regime is becoming more common in modern astronomy and should be considered useful instead of problematic, e.g., a notable example is that physical galaxy properties derived from SED fitting depend on priors and need a well-designed, physically motivated prior (e.g., \citealt{delaVega25}).

\bibliography{citations}{}
\bibliographystyle{aasjournalv7}
\end{document}